\documentclass[runningheads]{llncs}
\usepackage[height=235mm,width=155mm,center]{crop}
\usepackage[T1]{fontenc}
\usepackage{graphicx}

\usepackage{cite}
\usepackage{amsmath,amssymb,amsfonts}
\usepackage{algorithmic}
\usepackage{textcomp}
\usepackage{xcolor}

\usepackage{flushend} %
\usepackage{bbding}
\usepackage{wasysym}
\usepackage{pifont}

\usepackage{array}
\usepackage[normalem]{ulem}
\usepackage{booktabs}

\usepackage[hidelinks,implicit=true,bookmarksdepth=2]{hyperref}

\usepackage[a-1b]{pdfx}

\hypersetup{
	colorlinks=true,
	citecolor=blue,
	linkcolor=blue,
	urlcolor=blue,
}

\usepackage{pdfprivacy}

\usepackage{./orcidlink}

\renewcommand{\orcidID}[1]{\footnotesize\orcidlink{#1}\normalsize}

\usepackage{textpos}
\newcommand{\copyrightnotice}{
\begin{textblock}{9}(0.0,7.6)
	\noindent
	\fontsize{6pt}{6pt}\selectfont Paper published at UbiSec '23.\\ This version of the contribution has been accepted for publication, after peer review (when applicable) but is not the Version of Record and does not reflect post-acceptance improvements, or any corrections.\\The Version of Record is available online at: \url{http://dx.doi.org/10.1007/978-981-97-1274-8_26}.\newline Use of this Accepted Version is subject to the publisher’s Accepted Manuscript terms of use \linebreak \url{https://www.springernature.com/gp/open-research/policies/accepted-manuscript-terms}
\end{textblock}
\vspace{-1.3em}%
} %

\newcommand{\boldparagraph}[1]{\vspace{0.5em}\noindent\textbf{{#1}.}}

\usepackage{enumitem}
\newlist{RQLIST}{enumerate}{1}
\setlist[RQLIST]{label=\bfseries RQ\arabic*:, leftmargin=2.7em, parsep=0em}
 
\usepackage{pdfprivacy}

\begin{document}
\title{Is It Really You Who Forgot the Password? When Account Recovery Meets \\Risk-Based Authentication}
\titlerunning{Is it Really You Who Forgot the Password?}
\author{Andre Büttner \inst{1 (}\Envelope\inst{)}\orcidID{0000-0002-0138-366X} \and
Andreas Thue Pedersen\inst{1}\orcidID{0009-0009-8509-554X} \and
Stephan Wiefling\inst{2}\orcidID{0000-0001-7917-6065}
\and\\
Nils Gruschka\inst{1}\orcidID{0000-0001-7360-8314}
\and
Luigi Lo Iacono\inst{3}\orcidID{0000-0002-7863-0622}}
\authorrunning{A. Büttner et al.}
\institute{University of Oslo, Oslo, Norway \\ \email{\{andrbut,nilsgrus\}@ifi.uio.no} \and swiefling.de, Sankt Augustin, Germany
\\ \email{ubisec23@swiefling.de}\and
H-BRS University of Applied Sciences, Sankt Augustin, Germany \\ \email{luigi.lo\_iacono@h-brs.de}}
\maketitle              %
\copyrightnotice
\begin{abstract}
Risk-based authentication (RBA) is used in online services to protect user accounts from unauthorized takeover. RBA commonly uses contextual features that indicate a suspicious login attempt when the characteristic attributes of the login context deviate from known and thus expected values. Previous research on RBA and anomaly detection in authentication has mainly focused on the login process. 
However, recent attacks have revealed vulnerabilities in other parts of the authentication process, specifically in the account recovery function. Consequently, to ensure comprehensive authentication security, the use of anomaly detection in the context of account recovery must also be investigated. 

This paper presents the first study to investigate risk-based account recovery (RBAR) in the wild. We analyzed the adoption of RBAR by five prominent online services (that are known to use RBA). Our findings confirm the use of RBAR at Google, LinkedIn, and Amazon. Furthermore, we provide insights into the different RBAR mechanisms of these services and explore the impact of multi-factor authentication on them. Based on our findings, we create a first maturity model for RBAR challenges.
The goal of our work is to help developers, administrators, and policy-makers gain an initial understanding of RBAR and to encourage further research in this direction.

\end{abstract}

\section{Introduction}
Passwords are still the pre-dominant authentication method for online services, even for services that give access to confidential data or financial resources~\cite{quermann_state_2018,gavazzi_a_2023}. However, attacks on password authentication can be automated---e.g., credential stuffing using leaked passwords---and therefore scaled with little effort. This makes account takeover attacks on password-protected online services very lucrative for hackers~\cite{akamai_loyalty_2020}. As a countermeasure, more and more services offer multi-factor authentication (MFA) as an extension to password authentication. In this case, the user has to give additional proof of their identity, e.g., by entering a code from a one-time password (OTP) app or a text message (SMS). However, the additional step makes the authentication process more cumbersome and increases the risk of account lockouts in case the additional token gets lost~\cite{pohn2022multi}.

The idea of risk-based authentication (RBA)~\cite{freeman_who_2016,wiefling_is_2019,gavazzi_a_2023} is to balance security and usability. Here, the online service only requests additional authentication steps or blocks a client when it detects suspicious behavior. RBA does this by analyzing a set of feature values (e.g., location, browser, or login time) during the login process~\cite{gavazzi_a_2023,wiefling_is_2019}. 

A general problem with authentication is that the user might lose access to the authentication method---in the case of password authentication, this means primarily forgetting the password.
In such a case, the user has to pass the \textit{account recovery} process to regain access to their account. The process often involves sending a password reset link or an OTP to a pre-configured email address or phone number. If the required authentication (e.g., ownership of a phone, login to the email account) is weaker than the primary authentication, account recovery puts the overall account security at risk~\cite{cwe-recovery, microsoft_dev_2022}.

A high and common threat to account recovery mechanisms via email is when an attacker gains access to the corresponding email account, e.g., via credential stuffing~\cite{akamai_credential_2019,thomas_data_2017}. The recent FBI cybercrime report~\cite{federal_internet_2023} shows that compromised email addresses and phishing attacks are very popular attacks with potentially high financial loss for the hacked victims. 
Therefore, it is very important for online services to secure account recovery, for example, with MFA or RBA. So far, risk-based mechanisms have mostly been studied in the context of login authentication. 
However, we observed that mechanisms similar to RBA are also used for account recovery.

We define \textit{Risk-Based Account Recovery} (RBAR)\footnote{To the best of our knowledge, there is no standard term for it yet.} as a dynamic account recovery process on online services. It was indicated that such a method is used at a large online service~\cite{bonneau_secrets_2015}, but beyond that, RBAR and its appearances in the wild have not been publicly investigated yet. This is, however, important as it has the potential to protect a large number of users from account recovery attacks immediately. To learn about the current use of RBAR, we address the following research questions in this paper:

\begin{RQLIST}
    \item Do RBA-instrumented online services also use RBAR mechanisms? \label{rq1}
    \item What RBAR challenges are used in practice? \label{rq2}
    \item Are different RBAR challenges required when setting up MFA? \label{rq3}
\end{RQLIST}

\boldparagraph{Contributions}
This paper presents the first scientific insight into using RBAR in practice. We performed an exploratory analysis of RBAR behavior at Google and a systematic experiment on four other popular online services. We verified RBAR at three of the five services. The analysis also included the influence of MFA configurations and different (virtual) locations. The main contributions achieved from these activities are the following:

\begin{itemize}
    \item Identification of RBAR at popular online services
    \item A maturity model for different RBAR mechanisms
\end{itemize}
\vspace{1mm}
The remainder of this paper is structured as follows. Section \ref{sec:related_work} provides an overview of related work. In Section \ref{sec:rbar}, we describe details behind how RBAR works. Section \ref{sec:methodology} explains the methodology of our experiments. The findings of the two experiments are described in Sections \ref{sec:exploratory_study} and \ref{sec:experiment}, respectively. Our overall results are discussed in Section \ref{sec:discussion}. Section \ref{sec:conclusion} summarizes our work and suggests possible future work.

\section{Related Work}\label{sec:related_work}

Most of the previous work on account recovery considered it a static mechanism. 
For instance, a lot of research focused on different additional authentication challenges for recovery that can be solved easily by legitimate users but not by potential attackers. Examples include cryptographic keys~\cite{conners_lets_2019}, delegated account recovery~\cite{hill_moving_2017,javed_secure_2014}, dynamic security questions~\cite{addas_geographical_2019,hang_i_2015}, and email address or phone number verification~\cite{markert_work_2019}. While these works do not address risk-based use cases, we argue that such methods would be beneficial in conjunction with a risk analysis of the user context.

Further research evaluated online services in the wild. Li et al.~\cite{li_understanding_2020} studied the account recovery mechanisms of 239 popular online services in 2017 and 2019. They found that most of them implemented email address or mobile phone verification as a recovery mechanism. Amft et al.~\cite{amft2023lost} conducted a large-scale study investigating which recovery methods are usually deployed in conjunction with MFA methods. They unveiled that website documentation usually does not correspond with the actual recovery procedure, showing the lack of transparency in account recovery. We confirm this as we analyzed the documentation of the services we tested for any references to RBAR, which in most cases were absent (see Section \ref{sec:discussion}).

The only indication of risk-based recovery mechanisms we found in literature was mentioned by Bonneau et al.~\cite{bonneau_secrets_2015}, where they noted that Google performed a \textit{``risk analysis''} for account recovery. However, they did not further investigate how it works or what mechanisms are applied depending on the risk scenario.

Research on RBA is especially relevant for our work as it provides us with methods to analyze and develop risk-based systems.  For example, Wiefling et al.~\cite{wiefling_is_2019} studied RBA re-authentication mechanisms on five popular online services. They found that most online services used email verification to re-authen\-ticate users. Gavazzi et al.~\cite{gavazzi_a_2023} leaned on this work to identify that more than 75\% of the 208 studied online services do not use any form of RBA. While the research in this field only addresses plain user authentication, our work extends it by showing that the methods used in RBA research can be equally applied in the context of account recovery.
Consequently, we used the insights from prior work on RBA as a basis to study the use of RBAR on Google and other online services.

\section{Risk-Based Account Recovery}\label{sec:rbar}
\begin{figure}[t]
    \centering
    \includegraphics[width=\linewidth]{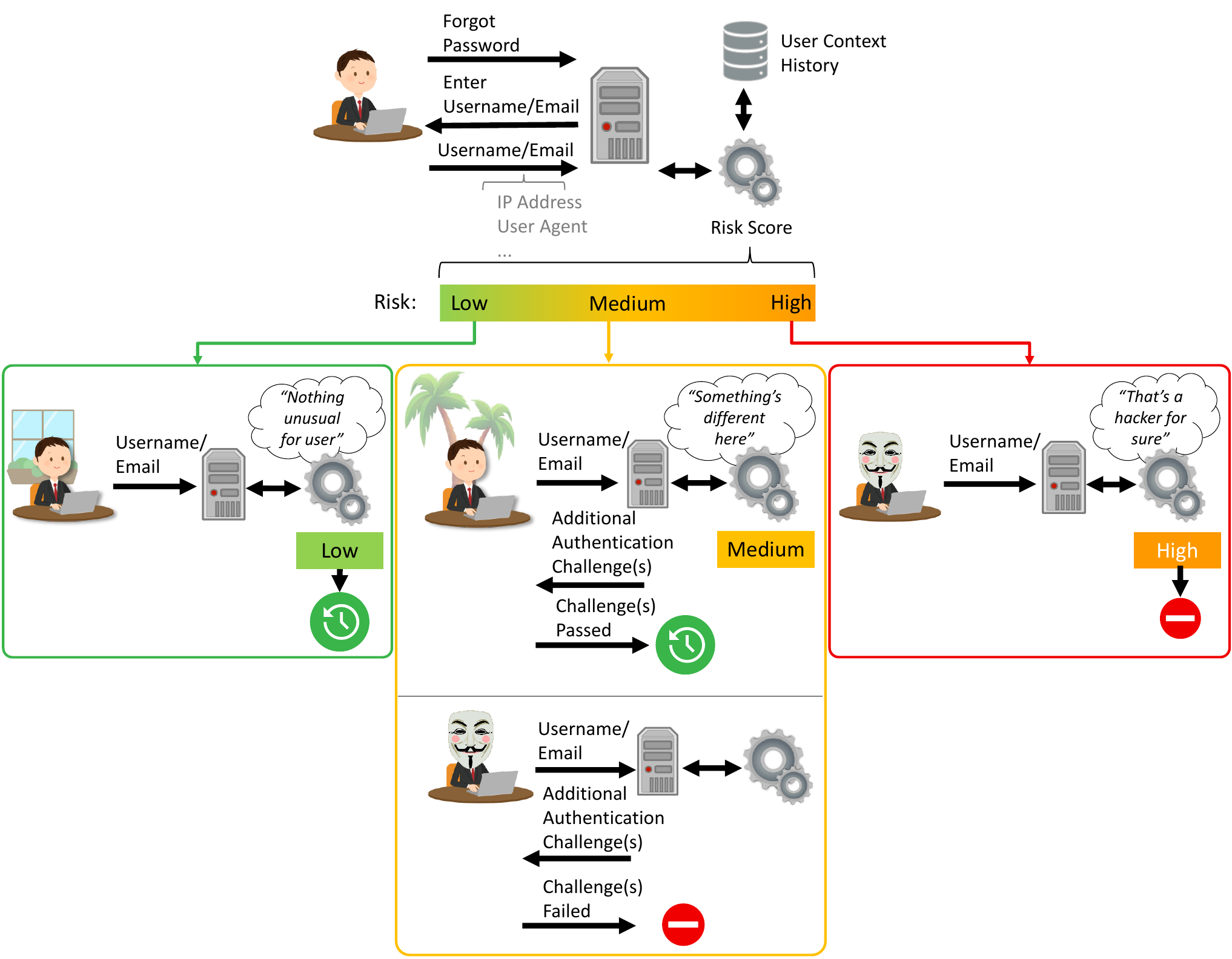}
    \caption{Overview of the RBAR procedure (based on RBA illustration in~\cite{Wiefling_Verify_2021})}
    \label{fig:rbar-illustration}
\end{figure}
Since there is no official description of RBAR yet, we describe its basic concept. Based on our observations on online services and previous knowledge in the related RBA field~\cite{wiefling_is_2019,Wiefling_Whats_2021}, RBAR works as follows (see Fig.~\ref{fig:rbar-illustration}):

A user typically starts an account recovery process, e.g., by clicking \textit{``forgot password''} at the online service's login form. After that, the user is asked to enter the username or email used for the account to recover. While submitting this identifier, the user also submits additional feature data that is available in the current context to the online service, e.g., IP address or user agent string. Based on this information, RBAR compares these values with the user context history and calculates a risk score. The user context history contains feature values of past user actions, like previous legitimate logins that might have been validated by RBA mechanisms~\cite{wiefling_pump_2023} before. The risk score is then classified into low, medium, and high risk. Based on the risk, the online service performs different actions.

At a \textit{low} risk, the feature values likely belong to the legitimate user, and the online service proceeds with the account recovery process (e.g., verify email address). A \textit{medium} risk occurs if the user's feature values deviate from the expected values. The online service then introduces additional authentication challenges that require more user effort (e.g., solving a CAPTCHA or answering questions related to the account). After successfully solving these challenges, the online service proceeds with the account recovery process.
A \textit{high} risk means that the online service suspects that the user is likely targeted by a hacking attempt. The online service might block the account recovery process in these cases. However, to avoid locking out legitimate users trying to recover their accounts, this possibility has to be carefully selected by the online service.

\section{Methodology}\label{sec:methodology}

We investigated the research questions by conducting two experiments. Prior research has indicated that Google applies risk-based decision-making for account recovery~\cite{bonneau_secrets_2015}, making it a suitable candidate for our first experiment. Therefore, we conducted an exploratory experiment on Google. We created test cases with different account setups, i.e., different authentication and recovery factor combinations. These were then tested with different user features to see how these could affect the recovery procedure. The study considered two RBA features, as suggested in Wiefling et al.~\cite{wiefling_is_2019}: known/unknown browser and known/unknown IP address. A \textit{known} browser is the one that was used before to sign in to Google, i.e., it has stored cookies from prior sessions. The \textit{unknown} browser was tested using the browser's incognito mode to have a clean browser session without previously set cookies. The IP address feature was varied by using a VPN connection to be able to study the uncertain area of medium to high risk scores~\cite{wiefling_is_2019}. By comparing the recovery procedures of the different features for each test case, we identified the mechanisms used for RBAR. The test cases and the final results are given in Section \ref{sec:exploratory_study}.

For the second experiment, we developed an improved and more systematic approach. As the experiment required manual effort, we limited the number of tested services to the following services that are known to use RBA~\cite{wiefling_is_2019,gavazzi_a_2023}:
\begin{itemize}
    \item LinkedIn (\texttt{linkedin.com})
    \item Amazon (\texttt{amazon.com})
    \item GOG (\texttt{gog.com})
    \item Dropbox (\texttt{dropbox.com})
\end{itemize}
The experiment was composed of three phases. First, we prepared user accounts for each service. Afterward, we checked whether any of the online services indicated RBAR behavior. Finally, since LinkedIn clearly turned out to implement RBAR, we analyzed if RBAR on LinkedIn is influenced by the MFA settings (as was the case with Google). More details on the steps and the results are presented in Section \ref{sec:experiment}. 

\section{Experiment 1: RBAR Use by Google}\label{sec:exploratory_study}

In the first experiment, we investigated previous assumptions~\cite{bonneau_secrets_2015} on whether Google used RBAR and identified features that might have an influence on the RBAR behavior. We describe the experiment and its results in the following.

\subsection{Preparation}\label{subsec:exp1_preparation}
The exploratory experiment on Google was conducted between October 2021 and March 2022. We set up four Google user accounts that were created at intervals of several weeks to mitigate being detected as a researcher. Based on the visible feedback from the online service, we assume that we remained under the respective detection thresholds.
In order to test the use of RBAR on Google, we defined the test cases based on the authentication and recovery factors offered in the Google account settings. At the time of the study, Google provided the following factors:
\begin{itemize}
\item \textbf{Main authentication}: password, sign in by phone
\item \textbf{Secondary authentication}: Google prompt, phone call or text message, backup codes, security key, authenticator app
\item \textbf{Recovery factors}: email, phone
\end{itemize}

The experiment on Google covered every possible single-factor authentication (SFA) account setup and eight MFA account setups. Each account setup was tested with all four RBA feature combinations. For each combination, all possible recovery options were explored.

\subsection{Results}
\begin{table}[t!]
    \caption{Examples for Google account recovery without MFA enabled}
    \label{tab:google_experiment_results_sfa}
    \centering
    \small
    \begin{tabular}{p{1.6cm} p{2.3cm} p{1.6cm} p{1.9cm} p{4.3cm}}
    \toprule
           \textbf{Recovery  \newline factor }  & \textbf{Phone signed \newline in} & \textbf{Known \newline browser} & \textbf{Known IP} & \textbf{Recovery procedure} \\%
           \midrule

        None & \Circle &  \CIRCLE & \CIRCLE & Recovery not possible\\\midrule%

         None & \CIRCLE & \CIRCLE & \CIRCLE & 1. Google prompt \\\midrule%

         None &  \CIRCLE & \Circle & \Circle & 1. Enter old password \newline 2. Google prompt (two steps) 
         \\\midrule%

        Email & \Circle & \CIRCLE & \CIRCLE & 1. Verify account email  \\\midrule%

        Email & \Circle &  \Circle & \CIRCLE &  1. Enter old password \newline 2. Verify account email  \\ \bottomrule
    \end{tabular}\mbox{}\\
    \vspace{0.2cm}
    \scriptsize
    \CIRCLE\textit{~=~Feature present, }\Circle\textit{~=~Feature not present}
\end{table}
The study found that Google used RBAR for both SFA and MFA account setups. This became clear as using an unknown browser and/or an unknown IP address increased the difficulty of recovering the account compared to using a known browser and IP address. This was indicated by requiring additional authentication factors, recovery options that were made unavailable, or an extra prompt like asking for the phone number of a registered phone.

\boldparagraph{Recovery Without MFA Enabled}
Table \ref{tab:google_experiment_results_sfa} lists a few examples\footnote{All results for the tests on Google are published on \url{https://github.com/AndreasTP/GoogleAccountRecovery}.} of the tests from studying SFA account recovery that clearly show the different recovery procedures based on RBA features. One can observe that in cases where an unknown browser was used for recovery, Google initially asked for an old password that the user could remember. This was not the case when using a known browser and a known IP address. The recovery procedure continued the same way, even if this step was skipped. 

When a phone was signed in to the same Google account, this phone was prompted with a button showing \textit{``Yes, it's me''}. Users had to click this button to confirm the ownership of the account. This behavior changed when trying to recover the account from an unknown browser and an unknown IP address. In this case, Google also showed a two-digit number on the recovery web page and presented a dialogue with three number options on the phone. Users then had to select the correct number on the phone to proceed with the recovery.

\boldparagraph{Recovery With MFA Enabled}
Table \ref{tab:google_experiment_results_mfa} shows some of the results that indicated obvious differences when trying to recover an account with a phone number configured for MFA. Note that in the given examples, we omitted the step of verifying access to the actual Google account email address to see what alternatives would be offered. When the recovery was performed from a known browser, it was sufficient to verify the phone that was set up for MFA by entering an OTP code that was sent to the phone via text message. Afterward, Google provided the user with an option to register and verify a new email address. A password reset email was sent to the newly registered email after 48 hours. In the meantime, the (legitimate) account owner got notifications about the ongoing recovery attempt. This allowed them to stop the procedure in case they did not request the recovery. However, this recovery option was not available when using an unknown browser. In that case, the user needed access to both the phone number and the email address registered on the actual Google account. This highlights how much RBAR features can impact the user's chance of a successful recovery.

\begin{table}[t!]
    \caption{Examples for Google account recovery with phone (text message) enabled for MFA}
    \label{tab:google_experiment_results_mfa}
    \centering
    \small%
    \begin{tabular}{p{2.8cm} p{2.8cm} p{1.9cm} p{4.3cm}}
    \toprule
        \textbf{Recovery factor}  & \textbf{Known browser} & \textbf{Known IP} & \textbf{Recovery procedure} \\
        \midrule
        None  & \CIRCLE & \CIRCLE\ / \Circle & 1. Verify MFA phone \newline 2. \sout{Verify account email}  \newline 3. Verify new email \newline $\rightarrow$ Reset email after 48hrs \\\midrule%

       None  &   \Circle & \CIRCLE & 1. Verify MFA phone  \newline 2. \sout{Verify account email} \newline $\rightarrow$ Recovery not possible \\\midrule%

        None &   \Circle & \Circle & 1. Enter MFA phone number \newline 2. Verify MFA phone  \newline 3. \sout{Verify account email} \newline $\rightarrow$ Recovery not possible \\
        \bottomrule
    \end{tabular}\mbox{}\\
    \vspace{0.2cm}
    \scriptsize
    \CIRCLE\textit{~=~Feature present, }\Circle\textit{~=~Feature not present, \sout{XXX}~=~Step omitted}
\end{table}

The last example in Table \ref{tab:google_experiment_results_mfa} shows a recovery procedure when using both an unknown browser and an unknown IP address. In this case, the user was first asked to enter the phone number used for MFA before actually verifying the ownership of this phone number. 

\begin{figure}
    \centering
    \includegraphics[width=0.6\linewidth]{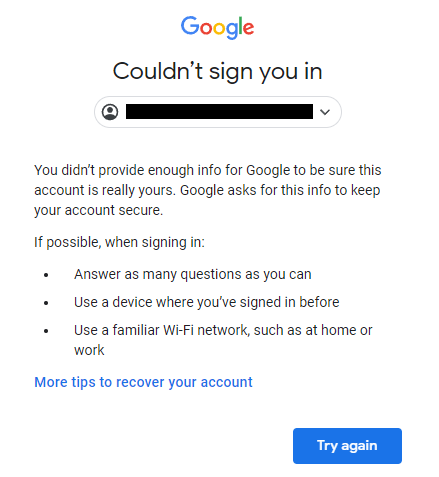}
    \caption{Message shown when failing Google's account recovery using an unknown browser and an unknown IP address. It reveals information that might give indications of their inner RBAR workings.}
    \label{fig:failed_recovery_message}
\end{figure}

\boldparagraph{Further Observations}
Also, we observed that when failing a recovery, Google revealed some information on how its RBAR mechanism might work. The message displayed to the user on a failed recovery attempt suggested using a known device and Wi-Fi during recovery (see Fig.~\ref{fig:failed_recovery_message}).

However, during the study, we experienced that the recovery process could change from one day to another. This was true despite using the same account, having the same recovery options configured, and using the same browser and IP address. For instance, a recovery procedure that earlier gave access to the account after 48 hours through a password reset email ended in a failed recovery. An authentication factor that could previously be used to help recover an account was occasionally removed as a recovery option. This suggests that Google uses more RBAR features than the two tested in this study.
Nonetheless, we confirm the assumption in prior work that Google implements a risk assessment in its recovery \cite{bonneau_secrets_2015}.

\section{Experiment 2: RBAR Use by Other Services}\label{sec:experiment}

The second experiment focused on online services that are known to use RBA~\cite{wiefling_is_2019} and investigated whether and how they also use some form of RBAR. We describe the experiment and its results below.

\subsection{Preparation}\label{subsec:preparation}
For this experiment, we began by setting up user accounts for all four online services (see Section~\ref{sec:methodology}).
Testing account recovery with personal accounts is not ideal since there is always the risk that accounts will be locked out or disabled entirely. However, RBA is oftentimes triggered only for legitimate accounts with a certain history of activity~\cite{wiefling_is_2019}. This makes sense from a technical perspective, as such algorithms need a certain amount of training data from the legitimate user to work correctly~\cite{wiefling_pump_2023}.
Therefore, we created four new test accounts for each of the services. These accounts were set up with the most basic settings, i.e., with a password and one email address. To avoid bias, we made sure to create and use new email addresses on general-purpose email providers not linked to universities for each account. In addition, we were able to provide one old account for each service, some of which were either personal or created in previous studies.

A training was conducted in which the test accounts were logged in more than 20 times within a time period of about 1.5 months (December 2022--January 2023). We based the number of logins on Wiefling et al.'s study~\cite{wiefling_is_2019}. Furthermore, it was ensured that the logins for each account were performed with a similar context, i.e., from the same browser and the same IP location. Also, logins from university IP addresses were avoided since experience has shown that online services might recognize these IP addresses and block accounts to prevent systematic analyses of their services. 
For reproducibility, we documented the context before each account login. We did this by recording all information from the IP address and HTTP header and the browser's internal JavaScript functions, as in related work~\cite{Wiefling_Whats_2021}.

\newcommand{\seq}{ $\rightarrow$ }
\newcolumntype{R}[1]{>{\raggedleft\let\newline\\\arraybackslash\hspace{0pt}}m{#1}}

\subsection{Identifying RBAR Usage}\label{subsec:discovery-rbar}
After training the test accounts, we analyzed whether the online services actually use RBAR mechanisms.
As in related work by Wiefling et al.~\cite{wiefling_is_2019} and Gavazzi et al.~\cite{gavazzi_a_2023}, this was tested by discovering differences in two distinct user contexts: \textit{normal} and \textit{suspicious}. This time, we considered a normal user to perform the account recovery from the same browser and IP location as in the training phase. In contrast, the suspicious user performs account recovery from a Tor browser. Web services can typically recognize Tor browser clients by the IP address of the exit nodes or by other browser features. Moreover, using a Tor browser is often considered suspicious~\cite{wiefling_is_2019}. We expected this to increase the likelihood of triggering risk-based mechanisms, if any, and compared to the first experiment on Google, where the Tor browser was not used. Note that we only considered differences that occurred after starting the recovery procedure for a specific account, e.g., after entering an email address. Any differences beforehand would not be relevant as it would mean that it is independent of the history of a user account.

\boldparagraph{Experimental Procedure}
For this within-group experiment, account recovery was performed twice for each test account on different days at the end of January 2023, once with a normal user context and once with a suspicious user context, in varying orders, to avoid bias. This means we performed two account recoveries with all provided accounts.
In the case of Amazon, we repeated the experiment with one of the new accounts and another old account due to inconsistent results, as described in more detail below.

\begin{table}[t!]
    \centering
    \caption{Account recovery procedures for a normal and suspicious user context for the different test accounts of each online service}
    \label{tab:study2-recovery-mechanisms}
    \small
    \begin{tabular}{p{2.9cm} l R{2.2cm} R{2.5cm}}
        \toprule
         \textbf{Online Service} & \textbf{Account} & \multicolumn{2}{l}{\textbf{User context}}\\
         && \multicolumn{1}{l}{Normal} & \multicolumn{1}{l}{Suspicious}
         \\
         \midrule
         Amazon & A1, A2, A4, A6$^\ast$ & EC  & EC\\
         &  A3, A1$^\dagger$ & CA\seq EC  & CA\seq EC \\
           & A5$^\ast$ & EC  & \underline{CA}\seq EC \\
          \midrule
         Dropbox & D1 -- D4, D5$^\ast$ & EL  & EL \\
          \midrule
         GOG & G1 -- G4, G5$^\ast$ & CA\seq EL  & CA\seq EL \\
          \midrule
         LinkedIn & L1 -- L4, L5$^\ast$ & EC  & \underline{CA}\seq EC \\
        \bottomrule
    \end{tabular}
    \mbox{}\\
    \vspace{0.2cm}
    \scriptsize
    \textit{EC = Email (Code), EL = Email (Link), CA = CAPTCHA, \\ $^\ast$ = Old account, $^\dagger$ = Experiment repeated, \underline{XXX} = Additional step}
\end{table}
\boldparagraph{Results}
Table \ref{tab:study2-recovery-mechanisms} summarizes the recovery procedures for each online service and account.
Overall, the presentation of a CAPTCHA was the only noticeable difference that was found. The CAPTCHAs in the table are underlined in those cases where they appeared only in the suspicious user context. Note that Amazon uses its own AWS WAF CAPTCHA~\cite{amazon-aws-captcha}, while GOG uses the Google reCAPTCHA v2~\cite{google-recaptchav2} and LinkedIn appears to use a custom CAPTCHA implementation. Dropbox did not use any CAPTCHA within our experiments.

For \textbf{Amazon}, in three cases, only an OTP code sent via email was requested. Afterward, the password could be changed. For one of the new test accounts~(A3), Amazon first requested a CAPTCHA before the email OTP code, but for both normal and suspicious contexts. For the old account~(A5), there was an actual difference as the CAPTCHA was only displayed in the suspicious context. Because of this inconsistent behavior, we did an additional test with A1, which this time required solving a CAPTCHA for both user contexts, similar to A3. Furthermore, we included a test with another personal account (A6) that was actively used to check if the behavior was related to the account age or activity. This time, no CAPTCHA had to be solved. Consequently, the risk assessment was more complex and could not be easily reproduced with our experimental setup.

\textbf{Dropbox} only requested the verification of the email address through a link before the password could be changed. This was the same for all user accounts, including the old one, and for both user contexts.

For \textbf{GOG}, a CAPTCHA had to be solved before verifying the email address through a link and finally changing the password. This was again equal for all accounts and both normal and suspicious user contexts.

\textbf{LinkedIn} was the only online service that consistently showed a different behavior depending on the context. For a normal user context, the email address had to be verified by an OTP code before the password could be changed. However, when performing recovery from a suspicious user context, a CAPTCHA had to be solved, sometimes multiple times.

In summary, Amazon and LinkedIn used RBAR, while Dropbox and GOG have not indicated any risk-based behavior during recovery. The only challenge that was shown depending on the user context was a CAPTCHA. The results for Amazon, however, were inconsistent for the different accounts. It was decided not to do a deeper analysis here, as the experimental setup clearly did not consider enough context parameters to simulate both a normal and a suspicious user context reliably. Yet, we conclude that Amazon must have used some form of RBAR. For LinkedIn, on the other hand, the RBAR behavior could clearly be reproduced with all accounts. Thus, we conducted a second experiment on LinkedIn using the newly created test accounts, as described in the subsequent section.

\begin{table*}[t]
    \centering
    \small
    \caption{Account recovery procedures for a normal and suspicious user context for the different LinkedIn account setups
    }
    \label{tab:study3-linkedin}
    \begin{tabular}{p{0.7cm} p{1.1cm} p{1.1cm} p{1.1cm} p{1.3cm} l l}
        \toprule
         \textbf{\#} & \multicolumn{2}{l}{\textbf{Recovery}}  & \multicolumn{2}{l}{\textbf{MFA}} & \multicolumn{2}{l}{\textbf{User context}} \\ %
         & Second Email & Text (SMS) & Auth. App & Text (SMS) & 
 \multicolumn{1}{l}{Normal} & \multicolumn{1}{l}{Suspicious}\\
         \midrule
         1 & \CIRCLE & \Circle  & \Circle & \Circle  &  EC1 $|$ EC2 & \underline{CA}\seq EC1 $|$ EC2 \\
         2 & \Circle & \CIRCLE & \Circle  & \Circle  & EC1 $|$ P1  & \underline{CA}\seq EC1 $|$ P1 \\
         3 & \Circle & \Circle & \CIRCLE  & \Circle &  EC1\seq AU  &  \underline{CA}\seq EC1\seq AU \\
         4  & \Circle  & \Circle & \Circle  & \CIRCLE  &  EC1\seq P2 &  \underline{CA}\seq EC1\seq P2 \\
         5 & \CIRCLE & \Circle  & \CIRCLE & \Circle &  EC1 $|$ EC2\seq AU  &  \underline{CA}\seq EC1 $|$ EC2\seq AU \\
         6 & \CIRCLE & \CIRCLE  & \Circle  & \CIRCLE &  EC1 $|$ EC2\seq P2   &  \underline{CA}\seq EC1 $|$ EC2\seq P2 \\
         7 & \Circle & \CIRCLE  & \Circle  & \CIRCLE &  EC1\seq P2  &  \underline{CA}\seq EC1\seq P2  \\
         8 & \Circle & \CIRCLE  & \CIRCLE  & \Circle &  EC1 $|$ P1\seq AU  &  \underline{CA}\seq EC1 $|$ P1\seq AU \\
         \bottomrule
    \end{tabular}%
    \small
    \mbox{}\\
    \vspace{0.2cm}
    \scriptsize
    \CIRCLE\textit{~=~Feature present, }\Circle\textit{~=~Feature not present, EC1 = Primary Email (Code), \\EC2 = Secondary Email (Code),  P1: Recovery Phone (SMS Code), \\P2 = MFA Phone (SMS Code), AU = Authenticator App, CA = CAPTCHA,}\\\ $|$~\textit{=~Alternative \underline{XXX} = Additional step
    }
\end{table*}

\subsection{Analyzing the Influence of MFA Settings on Account Recovery on LinkedIn}\label{subsec:influence-mfa}
In Section \ref{sec:exploratory_study}, we showed that Google implements RBAR by incorporating different authentication mechanisms that are set up as MFA factors in a user account. Since we could prove that LinkedIn also provides some form of RBAR, we conducted another experiment to determine whether LinkedIn used any other RBAR challenges beyond the CAPTCHA. 

\boldparagraph{Experimental Procedure}
For this experiment, we changed the authentication and recovery options in the LinkedIn test accounts. 
At the time of this experiment (January--February 2023), LinkedIn provided the following authentication and recovery methods:
\begin{itemize}
    \item \textbf{Main authentication}: password
    \item \textbf{Secondary authentication}: phone (SMS), authenticator app
    \item \textbf{Recovery factors}: email address, phone (SMS)
\end{itemize}
We tested the effects of all possible combinations of these methods. In addition, LinkedIn also offered a non-digital recovery method requiring the user to submit a copy of a government-issued ID. As this would have revealed the experimenters' identities, we did not include this method in the experiment. Similar to Google, the expected outcome for LinkedIn was that different authentication factors would be requested in a suspicious user context.

\begin{figure}[t!]
    \centering
\includegraphics[width=0.5\linewidth]{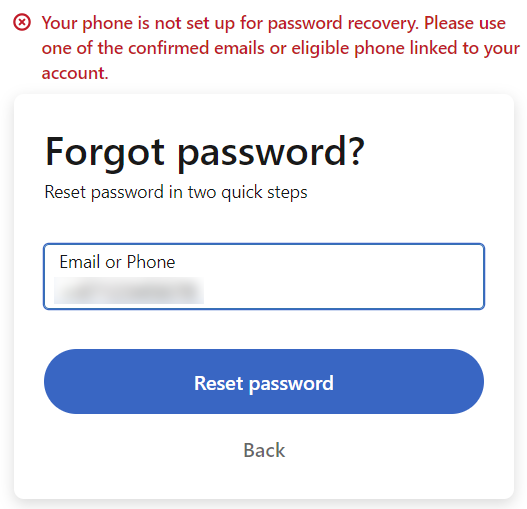}
    \caption{Error message for phone recovery, when also Text Message MFA is activated}
    \label{fig:linkedin-error}
\end{figure}

\boldparagraph{Results}
Table \ref{tab:study3-linkedin} shows the results for the different tested account setups. Note that in setups 1, 2, 5, 6, and 8, there are two possibilities for receiving the verification code: as an alternative to the primary email address, the secondary email address or the phone number could be entered (indicated by the ``$|$'' symbol). LinkedIn allows configuring the same phone number as a second authentication factor and as a recovery method. In fact, when enabling the phone number for MFA, the same number is activated automatically for recovery by phone. However, in such cases, using the phone for account recovery does not make much sense as only a single factor (ownership of the SIM card) is required for resetting the password and logging in afterward, which contradicts the idea of \textit{multi}-factor authentication. In these cases, i.e., setups 6 and 7, we only received an inaccurate error message (see Fig.~\ref{fig:linkedin-error}). We filed a bug report for this to LinkedIn on February 24, 2023. However, the response from LinkedIn (one day later) indicated that it will not be fixed anytime soon unless it gets noticed by several other users.

The experiments show that the behavior when configuring further recovery or authentication methods is identical to the base setup. The only difference in the account recovery procedure for all setups was the initial CAPTCHA shown in the suspicious user context. Apart from that, the account recovery procedure always started with the verification of the primary email address or phone number by an OTP code, followed by the verification of the MFA method if one was activated.

\boldparagraph{Variation of CAPTCHA Iterations}
\begin{table}[t!]
    \centering
    \small
    \caption{Number of CAPTCHA iterations for different (pretended) locations for account recovery on LinkedIn}
    \label{tab:linkedin-captcha}
    \begin{tabular}{p{4cm} l}
        \toprule
            \textbf{CAPTCHA iterations}  & \textbf{Location of Tor exit} \\
        \midrule
        1 & Sweden, Poland, United Kingdom, \textit{Mexico} \\
        2 & United Kingdom, Germany \\ 
        3 & 3$\times$ USA, \textit{Czech Republic} \\
        5 & USA, Canada, \textit{Netherlands} \\
        \bottomrule
    \end{tabular}
\end{table}
In addition to our main results, we observed that the number of iterations of the CAPTCHA on LinkedIn varied in different experiments between 1 and 5. When mapping the number of iterations to the pretended location (i.e., the location of the Tor exit node), an interesting correlation showed up (see Table~\ref{tab:linkedin-captcha}). The normal usage location for all accounts was in Europe, and when the pretended location was also in Europe (just another country), 1 or 2 repetitions of the CAPTCHA were required. In cases where the suspicious location was on a different continent, 3 or 5 repetitions were needed. However, there were also cases (marked in italics) where this was not true. Nonetheless, it indicates that LinkedIn's RBAR might give different suspicious risk classifications that are reflected in the number of CAPTCHA iterations. It also seems that the location is one important feature. Further experiments are needed to analyze to what extent other features are included.

\section{Results and Discussion}\label{sec:discussion}
In our exploratory study on Google and the follow-up experiment with other online services, we confirmed that several online services apply RBAR to a certain degree. 
In this section, we describe the results of the experiments with regard to the research questions. Furthermore, we summarize the results in a maturity model that we propose for RBAR implementations. Finally, we outline the limitations of our experiments and discuss further aspects of RBAR usage in practice.

\subsection{Experiment Results}

Within the scope of our experiments, we observed that Google implements RBAR in quite a sophisticated manner. It showed different authentication methods depending on the account setup and the user context. Dropbox and GOG did not apply any risk-based mechanisms during account recovery. Amazon actually indicated the use of RBAR, however, by assessing context information that was not considered by our two different user contexts. In some tests, a CAPTCHA had to be solved, while in others, it was not required. LinkedIn clearly behaved differently in a suspicious user context. When trying to recover an account from a Tor browser, LinkedIn showed a CAPTCHA challenge before entering an email verification code. In contrast to Google, however, the RBAR for LinkedIn did not involve MFA settings in a user account.

With regard to \textbf{RQ1}, we conclude that there are online services that use RBA, which also use RBAR---including Google, Amazon, and LinkedIn---but not all of them. 
To answer \textbf{RQ2}, the challenges we found on Google include pre-configured MFA methods (e.g., phone number) and questions requiring background knowledge (e.g., old passwords). On LinkedIn and Amazon, we only observed a CAPTCHA challenge in connection with RBAR. Concerning \textbf{RQ3}, we found that the MFA settings influenced the recovery procedure on Google only, while LinkedIn did not vary RBAR challenges depending on any configured MFA methods.

\subsection{Maturity Model}

\begin{table*}[t!]
\caption{Maturity model with maturity levels, mapping of RBAR challenges to the tested services and possible attacks against these challenges}
    \label{tab:results_overview}
    \centering
    \small
    \begin{tabular}{p{1.6cm}p{3.25cm}p{2.65cm}p{4.3cm}}
         \toprule
            \textbf{Maturity }&\textbf{RBAR challenge} & \textbf{Identified on} &  \textbf{Possible attacks }\\
        \midrule
        3 & Pre-configured MFA & Google  & Physical attack, malware \cite{campobasso_impersonation_2020} \\
        2 & Background knowledge & Google & OSINT, leaked passwords, phishing \cite{hang_i_2015,addas_geographical_2019}  \\
        1 & CAPTCHA  & LinkedIn, Amazon &  Manual recovery, CAPTCHA bypass algorithm \cite{hossen2020object,sukhani2021automating} \\
        0 & None & Dropbox, GOG & n/a\\
        \bottomrule
    \end{tabular}
\end{table*}
Based on our results and inspired by \cite{pohn2022multi}, we propose a maturity model that ranks the different RBAR challenges by difficulty for an attacker (see Table~\ref{tab:results_overview}). Due to the nature of RBAR, the model only considers the measures used in connection with a risk assessment. It describes the additional security gain in case the primary recovery factor (e.g. email address), if any, has already been compromised. Thus, no RBAR at all is considered the least mature as it does not involve any risk assessment and does not provide additional measures. Showing a CAPTCHA is ranked as level 1 as it can prevent automated attacks. Yet an attacker might bypass it or manually exploit account recovery. Background questions are ranked as level 2 as they require an attacker to gather knowledge of a victim. However, it also increases only the cost of the attack. MFA methods that are pre-configured in an account are considered the most mature as they require more sophisticated methods or even physical access for a successful attack. 

The model can be used, e.g., to assess the security of an RBAR implementation. Online services can also use such a model for their RBAR implementations to enable certain challenges with a higher maturity ranking at higher risk scores. Note that the model is only one possible way to assess RBAR. It might be different if other types of RBAR challenges are used that were not discovered within our study.

\subsection{Comparison with Official Documentation}\label{sec:comparison-with-documentation}

To the best of our knowledge, the experiments showed for the first time that Amazon, LinkedIn, and Google use RBAR. To compare our findings with the public communications of the online services, we took their official documentation into consideration~\cite{amazon_reset_2021,dropbox_how_2023,gog_how_2023,google_tips_2023,linkedin_password_2022}. Interestingly, none of the RBAR-instrumented online services mentioned that they change the account recovery behavior based on contextual information collected during the recovery process~\cite{amazon_reset_2021,linkedin_password_2022,google_tips_2023}. Only Google hinted that users should possibly use a familiar device and location~\cite{google_tips_2023}. However, they did not mention why users should do this, i.e. because they use RBAR. Our results show that the account recovery mechanisms of these online services seem to do more to protect their users than what is officially communicated to them.

Trying to hide implemented security mechanisms from the user base has already been observed in the related case of RBA~\cite{golla_i_2019} and other research on account recovery~\cite{amft2023lost}. We do not consider this a good practice, as it follows the anti-pattern of \textit{``security by obscurity''}. Users also tend to get frustrated when they experience security barriers that were not communicated to them beforehand~\cite{Wiefling_More_2020}. Beyond that, attackers are known to adapt to obscured security mechanisms~\cite{milka_anatomy_2018,thomas_data_2017}. We assume that public RBAR research, to increase the body of knowledge, will increase the overall adoption of online services and enable a large user base to be protected with RBAR following the principle of \textit{``good security now''}~\cite{garfinkel_design_2005}.

\subsection{Ethics}
We only tested account recoveries with accounts owned by the researchers, i.e., we did not try to exploit the recovery of other users' accounts. Also, since we conducted manual tests, we did not create high traffic on the online services that could have affected other users.

While it could be reasoned that our findings are helpful for attackers, we argue that they are more valuable to the public. As the gained knowledge helps researchers and online service providers to get an understanding of how RBAR works, this can support the development of more secure and usable account recovery mechanisms.

\subsection{Limitations}
Beyond Google, only four online services were analyzed in terms of RBAR. This was mainly due to the lack of any automatism for training user accounts and testing account recovery, therefore requiring manual effort to conduct our experiments. Nevertheless, as mentioned before, these services have been carefully selected as they are known to use RBA~\cite{wiefling_is_2019}.

We could not find any RBAR mechanisms in Dropbox and GOG. Due to the nature of a black-box test, we do not know the implementation details of the tested online services. Thus, there is always uncertainty involved. Nevertheless, we are confident that the accounts were sufficiently trained---especially since we also tested older accounts---and tested with the highest risk possible~\cite{wiefling_is_2019}.

\subsection{RBAR}
Attackers may abuse account recovery to circumvent authentication. Hence, the security of account recovery is as essential as the security of login authentication. Previous research showed that email addresses often become a single point of failure~\cite{li2018email,li_understanding_2020}. RBAR might be an advantageous way to increase the difficulty of a successful account takeover by incorporating additional authentication methods, as with RBA. At the same time, it may reduce the burden on legitimate users and increase their chances of recovering an account. 

The RBAR used by Google is quite different from LinkedIn. Google uses additional authentication methods, while LinkedIn just requires a suspicious user to solve an additional CAPTCHA. This CAPTCHA actually only reduces the risk of automated attacks by making it more costly for an attacker. In general, CAPTCHAs mainly increase friction for users \cite{yan2008usability}. It may be an improvement to use a risk score to decide if a CAPTCHA should be solved, compared to, e.g., GOG, where a CAPTCHA is shown to all users. However, the security gain is insignificant since researchers have already demonstrated attacks against Google's widely known reCAPTCHA~\cite{hossen2020object,sukhani2021automating}. Moreover, this does not prevent targeted attacks. We argue that if a service already implements a risk assessment in its account recovery, it should even go further and include actual authentication methods. In the case of LinkedIn, it could, for instance, request the verification of another recovery email or phone if set up.

\section{Conclusion}\label{sec:conclusion}
Account recovery mechanisms remain a relevant entry point for account takeover attacks~\cite{cwe-recovery, microsoft_dev_2022}. Online services should strengthen their account recovery with additional security mechanisms, like risk-based account recovery (RBAR), to protect their users.

In this paper, we investigated the use of RBAR in practice. We described the concept behind RBAR and conducted two experiments to learn about if and how online services use it. The results show that Google, Amazon and LinkedIn used RBAR. However, their implementations differed widely in suspicious contexts, from asking users for background knowledge or pre-configured MFA methods (Google) to showing a CAPTCHA challenge (Amazon and LinkedIn). Based on our results, we proposed a maturity model that researchers or service providers can use to assess the security of RBAR systems or guide in implementing RBAR.

Following this first systematic analysis of RBAR, future work can extend our proposed model with other RBAR challenges.  Furthermore, it can be studied what features specifically trigger RBAR challenges. As there seems to be a tendency to include risk-based decision-making into account recovery, there should be a comparison of RBA and RBAR and how they can complement each other in authentication systems as a whole.
 \subsubsection{Acknowledgments} 
Stephan Wiefling did this research while working at H-BRS~University of Applied Sciences.
\bibliographystyle{splncs04}
 \bibliography{references}
\end{document}